\documentclass[10pt,twocolumn,superscriptaddress,english,pra,showpacs,floatfix,aps]{revtex4-2}
\usepackage[utf8]{inputenc}
\usepackage{amsmath}
\usepackage{amssymb}
\usepackage{graphicx}
\usepackage{xspace}
\usepackage{mathtools}
\usepackage{comment}
\usepackage{physics}
\usepackage{bm}
\usepackage{babel}

\makeatletter
 
\usepackage[normalem]{ulem}
\usepackage{xcolor}
\usepackage{soul}
\usepackage[backref=none,
bookmarksnumbered=true,
bookmarks=true,
bookmarksopen=true,
colorlinks=true,
citecolor=blue,
linkcolor=blue,
anchorcolor=green,
urlcolor=blue,unicode=false]{hyperref}

\makeatother

\begin{document}

\preprint{APS/123-QED}

\title{Multi-band fractional Thouless pumps}

\author{Marius J\"urgensen}
    \email{marius.juergensen@gmail.com}
    \affiliation{Department of Physics, The Pennsylvania State University, University Park, Pennsylvania 16802, USA}
    \affiliation{Department of Physics, Stanford University, Stanford, California 94305, USA}%
\author{Jacob Steiner}
    \affiliation{Department of Physics and Institute for Quantum Information and Matter, California Institute of Technology, Pasadena, California 91125, USA}%
\author{Gil Refael}
    \affiliation{Department of Physics and Institute for Quantum Information and Matter, California Institute of Technology, Pasadena, California 91125, USA}%
\author{Mikael C. Rechtsman}%
    \affiliation{Department of Physics, The Pennsylvania State University, University Park, Pennsylvania 16802, USA}%

\date{\today}

\begin{abstract}

Quantization of particle transport lies at the heart of topological physics. In Thouless pumps -- dimensionally reduced versions of the integer quantum Hall effect -- quantization is dictated by the integer winding of single-band Wannier states. Here, we show that repulsive interactions can drive a transition from an integer- to a fractional-quantized Thouless pump (at fixed integer filling) by stabilizing a crystal of multi-band Wannier states, each with fractional winding. We numerically illustrate the concept in few-particle systems, and show that a dynamical Hartree-Fock ansatz can quantitatively reproduce the pumping phase diagram. 

\end{abstract}

\maketitle

Thouless pumps \cite{thoulessQuantizationParticleTransport1983} are one-dimensional time-periodic systems that inherit their topological properties from two-dimensional Chern insulators via dimensional reduction, i.e. by replacing one wavevector dimension with a time-periodic modulation. Topology manifests as quantization of particle transport: provided the modulation is adiabatic, an integer number of particles per pump cycle is transported past any point, given by the Chern numbers of occupied bands \cite{thoulessQuantizationParticleTransport1983}. Thouless pumps have been realized in a variety of platforms \cite{citroThoulessPumpingTopology2023}, including fermionic and bosonic ultra-cold atoms in lattices \cite{lohseThoulessQuantumPump2016,nakajimaTopologicalThoulessPumping2016,luGeometrical2016,dreon2022self}, mechanical systems \cite{grinberg2020robust,XiaExperimental2021}, plasmonics \cite{fedorova2020observation}, and photonic waveguide arrays \cite{krausTopologicalStatesAdiabatic2012,ke2016topological,cerjan2020thouless}. Recently, it has been suggested that topological pumping may find use in quantum information processing, enabling robust state-independent optical transport \cite{zhu2024quantum}.

Early on, Niu and Thouless demonstrated that the integer quantization of pumped electronic charge is robust to interactions as long as the ground state is unique, and the system remains gapped throughout the pump cycle \cite{niuQuantisedAdiabaticCharge1984,haywardTopologicalChargePumping2018}. Recent experiments have sought to overcome these limitations and investigated the breakdown of quantization due to interactions \cite{walterQuantizationItsBreakdown2023}, as well as interaction-induced quantization \cite{viebahnInteractions2024,luengoStabilization2024}. While non-interacting Thouless pumps are well understood in terms of their analogy to the integer quantum Hall effect, the understanding of interacting Thouless pumps, and their possible connection to the fractional quantum Hall effect \cite{tsuiTwoDimensionalMagnetotransportExtreme1982}, is still developing. 

Previous studies \cite{grusdtRealizationFractionalChern2014, zengFractionalChargePumping2016,taddiaTopologicalFractionalPumping2017} have shown that in interacting Thouless pumps with repulsive interactions, fractional filling can lead to fractional pumping.  This arises from charge density wave formation, which effectively increases the lattice periodicity. However, for weak interactions, such systems do not exhibit quantized pumping. Fractional pumping of a distinctly different nature has also been observed in non-interacting systems with additional symmetries in the synthetic Brillouin zone, occurring at fractional periods of the pumping cycle \cite{marra2015fractional}.

Here, we propose a Thouless pump exhibiting both integer and fractional pumping at fixed integer filling controlled by interaction strength. We were motivated by recent experiments in photonic waveguide arrays that exhibit fractional pumping of solitons \cite{jurgensenQuantizedNonlinearThouless2021,jurgensenQuantizedFractionalThouless2023}, in which the fractional displacement per period is due to the fractional winding of multi-band Wannier states \cite{vanderbiltBerryPhasesElectronic2018}. We introduce a fermionic model that exhibits fractional pumping based on a similar mechanism: spontaneous polarization of multi-band Wannier states. Key ingredients are (1) a set of energetically isolated topological bands, separated by small band gaps and with Chern numbers whose sum is not divisible by the number of bands in this set
and (2) strong finite-range repulsive interactions extending over multiple sites that couple these bands. These conditions may e.g. be realized in cold atomic systems through Rydberg interactions. Beyond a critical interaction strength, occupation of multi-band Wannier states is energetically favorable compared to the unpaired band counterparts due to the superior localization.

To illustrate the effect, we focus on the simplest realization, a pair of topological bands at unit filling. Using numerical techniques and a toy model, we show fractional pumping in an intermediate adiabatic regime without broken lattice translation symmetry.
For small systems, the transferred charge is fractionally quantized with corrections that are exponentially small in the ratio of interaction strength to total bandwith, even after many drive periods. We expect that in large systems approximate fractional pumping persists, up to a prethermal time scale. Finally, we show that the effect is well-captured by a mean-field approach, demonstrating the pumped charge transitions from integer at weak, to non-quantized at intermediate, to fractional at strong interactions.

\begin{figure}[ht]
    \includegraphics{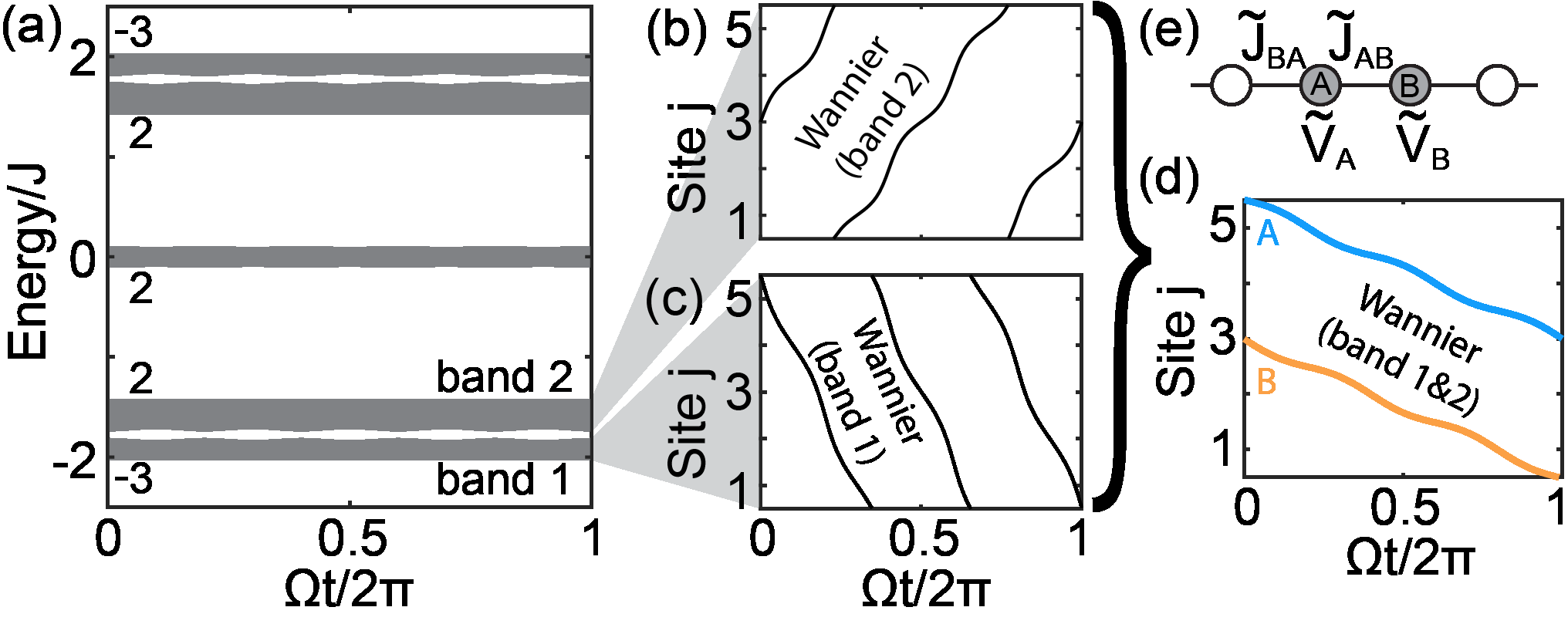}
    \caption{Properties of the noninteracting AAH-model. 
    (a) Band structure of the noninteracting AAH model with Chern numbers. Bands 1 and 2 are close in energy and well separated from all other bands. (b-c) Single-band Wannier centers of band 1 (c) and band 2 (b), whose number of windings equals the Chern number of the band. (d) Wannier centers of maximally-localized multi-band Wannier states of the lowest two bands. Note that there exist two per unit cell, denoted here by $A$ and $B$, each with fractional, but combined integer winding. (e) Effective two-site model for the lowest two bands, where each site represents a Wannier state. Parameters: $K/J = 0.7$, where $J$ is the nearest-neighbor hopping strength and $K$ the strength of the hopping modulation.}
    \label{ED-fig:model}
\end{figure}

\paragraph*{Model.} 
We consider an interacting, spinless, off-diagonal version of the Aubry-André-Harper (AAH) model \cite{harperSingleBandMotion1955,aubryAnalyticityBreakingAnderson1980} with Hamiltonian $H(t) = H_0(t) + H_{\text{int}}$: 
\begin{align}
    H_0(t) &= -\sum_{j} J_{j}(t)\hat{c}_{j+1}^{\dag}\hat{c}^{}_{j} + \textrm{h.c.}, \nonumber \\
    H_{\text{int}} &= \sum_{j,r\geq 1} U^{}_{r} \hat{n}_{j}\hat{n}_{j+r},
    \label{ED-eq-model}
\end{align}
where $\textrm{h.c.}$ denotes the Hermitian conjugate, $\hat{c}_{j}^{\dag}$ ($\hat{c}^{}_j$) is the fermionic creation (annihilation) operator for site $j$, $\hat{n}_j = \hat{c}_{j}^{\dag}\hat{c}^{}_j$ is the associated particle number operator, and $U_r$ describes the strength of interactions between fermions at nearest-neighbor ($r$=1), next-nearest-neighbor ($r$=2) sites and so on. The non-interacting Hamiltonian, $H_0(t)$, describes a Thouless pump;  we choose an AAH-model with five sites per unit cell and time-periodic nearest-neighbor hopping strengths $J_{j}(t) = J + K \cos(\Omega t+ 4\pi j/5)$. $\Omega=2\pi/T$ defines the frequency of the periodic modulation with period $T$ and goes to zero in the limit of adiabatic driving. We consider unit band filling: the number of particles equals the number of unit cells, $N$.

The behavior of Thouless pumps can be understood in terms of the instantaneous eigenstates of $H(t)$. We consider first the noninteracting system $H_0(t)$ with five bands as shown in Fig.~\ref{ED-fig:model}(a). The lowest two bands are close in energy but well separated from the rest, allowing us to focus on the former. Quantization of pumped charge is intimately related to the winding of Wannier states over the pump cycle. Specifically, the winding of the Wannier center trajectory is equal to the Chern number of the associated band, and hence the pumped charge. Figure ~\ref{ED-fig:model}(b,c) shows the real space trajectories of the Wannier centers of the lowest two bands, with Chern numbers of $C=-3$ and $C=+2$, respectively. A (single-band) Thouless pump thus can be intuitively understood as the Wannier state in each unit cell being occupied and displaced by the Chern number during one pump cycle.

\begin{figure}
    \centering    \includegraphics[width=\linewidth]{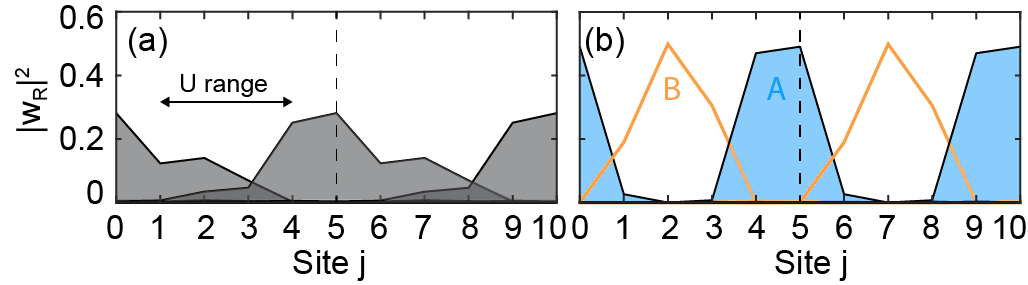}
    \caption{    
    Localization properties and interaction energy of maximally-localized Wannier states. Shown are the probability densities, $|w_R|^2$, of (a) single-band Wannier states of band 1, and (b) both multi-band Wannier states of band 1 and 2 from Fig.~\ref{ED-fig:model} for a representative time-point in the pumping cycle $\Omega t/2\pi = 0.37$. Note that in the multi-band case there exist two types of Wannier states per unit cell, labelled as $A$ and $B$. At integer filling, one Wannier state per unit cell is filled (shaded). The arrow indicates the range of interactions used in the simulations.}
    \label{fig:wannier_occupation}
\end{figure}

We now add repulsive interactions, strong enough to couple the low energy bands $1$ and $2$ but insufficient to involve higher bands. Heuristically, the interaction energy is minimized by maximally localizing the fermions given the available states. Starting from the filled first band, for which maximally-localized single-band Wannier states give the most localized basis, one may further reduce the interaction energy by mixing in states from the second band to further localize the states. We will see below that the interaction energy is minimized if all fermions occupy one flavor of maximally-localized multi-band Wannier states \cite{vanderbiltBerryPhasesElectronic2018}, 
\begin{equation}
    \ket{w_{R,\alpha}(t)} =\sum_{k,m\in\{1,2\}} \mathcal{U}_{m\alpha}(k,t)  \frac{e^{-ikR}}{\sqrt{N}} \ket{\phi_{k,m}(t)},
\end{equation}
where the unitary matrices $\mathcal{U}_{m\alpha}(k,t)$ determine the localization and $\ket{\phi_{k,m}(t)}$ are instantaneous Bloch states. There are two such Wannier states per unit cell, denoted by $\alpha \in \{A,B\}$, referred to as flavors here. Their degree of localization compared to the Wannier states of the first band is depicted in Fig.~\ref{fig:wannier_occupation}. While the exact form of the interaction is not crucial, it has to be of sufficient range to induce considerable repulsion between maximally-localized single-band Wannier states and multi-band Wannier states of different flavors. Here, we use up to next-next-nearest neighbor interactions (see Fig.~\ref{fig:wannier_occupation}).

\begin{figure*}[t]
    \includegraphics[width=\linewidth]{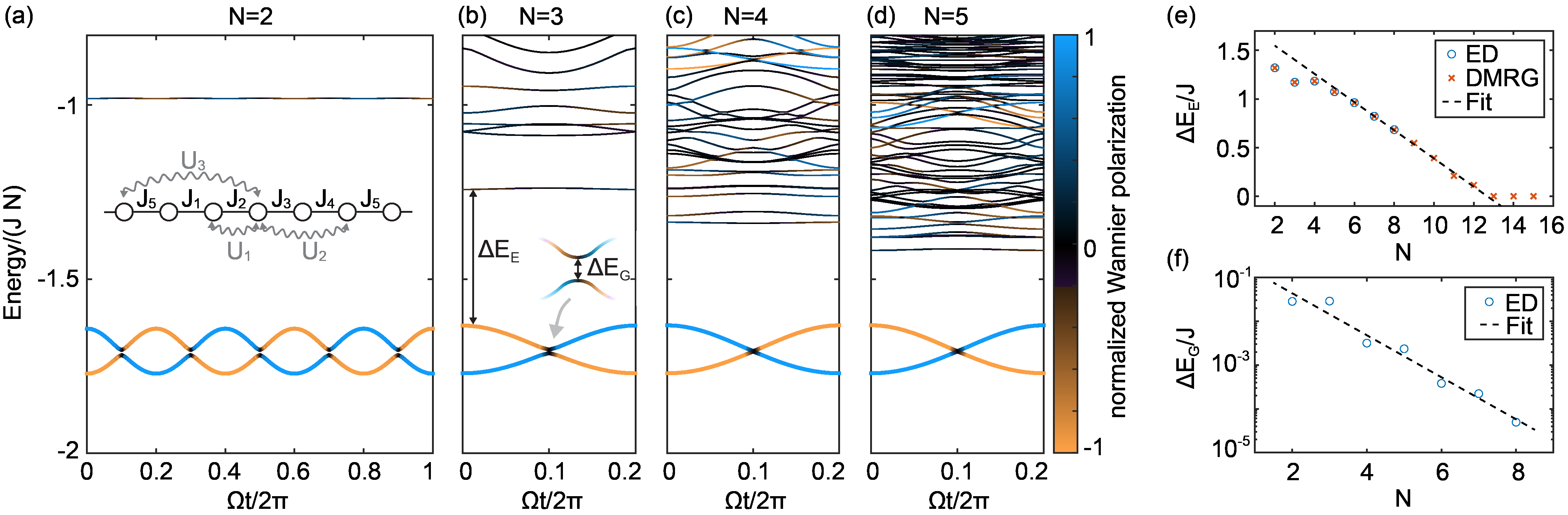}
    \caption{
    Properties of the interacting AAH-model. (a-d) Instantaneous many-body energies during the pumping cycle for $N=2,3,4$ and $5$ particles and $5N$ sites. Color coding indicates the (normalized) polarization of the states in terms of multi-band Wannier flavors. Blue (orange) denote that a state is composed of only one of the two flavors of multi-band Wannier states. Note that (b-d) only show one fifth of a pumping cycle. The inset in (a) shows a schematic of the interacting AAH model. The inset in (b) defines the energy gaps, that are further analyzed in (e) and (f). (e) Linearly decreasing energy gap between the ground state manifold and higher energy states, $\Delta E_E$, as a function of system size $N$. Circles represent ED results and crosses represent DMRG calculations. (f) Exponentially decreasing energy gap between the polarized low-energy states, $\Delta E_G$, as a function of $N$. Parameters are identical to Fig.~\ref{ED-fig:model}, but with additional interaction strength $U_r$ between neighbors of distance $r$: $U_1/J = 40$, $U_2/J = 10$, $U_3/J = 3$.}
    \label{ED-fig:Fig2} 
\end{figure*}

Unlike single-band Wannier centers, multi-band Wannier centers do not need to return to themselves after one period, but can instead connect to a different multi-band Wannier center, so that their winding is only defined after multiple periods. We refer to this as fractional winding. 
Figure~\ref{ED-fig:model}(d) shows the trajectories for the maximally-localized multi-band Wannier states of the lowest two bands. Each state shows fractional winding with interchanging flavors. However, together both multi-band Wannier states wind by $-1$, which equals the sum of the Chern numbers $C_{1,2} = -3 + 2 = -1$.

We arrive at the following picture: strong interactions favor a state that occupies one flavor of multi-band Wannier state, say $\ket{\Psi_B(t)} = \prod_R \hat{w}^\dagger_{R,B}(t) \ket{\emptyset}$, where $\hat{w}^\dagger_{R,\alpha}(t)$ creates $\ket{w_{R,\alpha}(t)}$ and $\ket{\emptyset}$ is the vacuum. Initializing the system in this state, and assuming that the occupations do not change throughout the pump cycle, after one period the state is $\ket{\Psi_B (t+T)}  = \ket{\Psi_A(t)} = \prod_R \hat{w}^\dagger_{R,A} (t)\ket{\emptyset}$. The charge pumped up until this point is not quantized. However, after a second period, the state returns to $\ket{\Psi_B(t)}$, and all occupied Wannier states shifted by $C_{1,2} = -1$. This constitutes a fractional Thouless pump: the time-averaged transferred charge per period is $-1/2$. In the following, we will verify this picture numerically and within a mean-field approach, and establish the parameter regimes in which fractional pumping occurs.

We note that the proposed mechanism is more general than the two-band situation considered here. In Thouless pumps with $m$ nearly degenerate bands the winding of multi-band Wannier states may require up to $m$ periods. At single-band filling, this implies a pumped charge per period equal to the sum of the Chern numbers divided by $m$, provided interactions stabilize a multi-band Wannier ground state.

\paragraph*{Numerical results.}

Using exact diagonalization, we compute the instantaneous many-body spectra of Eq.~\eqref{ED-eq-model} as a function of $\Omega t$, and for $N=2,3,4$ and $5$ particles, see Figs.~\ref{ED-fig:Fig2}(a-d). The states are colored by their multi-band Wannier polarization, $\sum_R \bra{\Psi(t)} \hat{n}_{R,A}(t) - \hat{n}_{R,B} (t)\ket{\Psi(t)}$, normalized by $\sum_R \bra{\Psi(t)} \hat{n}_{R,A}(t) + \hat{n}_{R,B}(t) \ket{\Psi(t)}$, with $\hat{n}_{R,\alpha}(t) = \hat{w}^\dagger_{R,\alpha}\hat{w}^{}_{R,\alpha}(t)$ and $\ket{\Psi(t)}$ the respective instantaneous many-body eigenstate.
Two nearly fully polarized low-energy states emerge, corresponding to $\ket{\Psi_A(t)}$ and $\ket{\Psi_B(t)}$. They exhibit an odd number of anti-crossings throughout the pump cycle, reflecting the interchange of Wannier flavors after one period (inverted coloring at $t=T$ vs. $t=0$). A weak perturbation can remove all but one of these crossings, c.f. the supplemental material \cite{SI}. Fractional pumping occurs when the state of a given flavor is followed. This requires $\Omega \gg \Delta E_G$, the gap at the anti-crossings, while remaining adiabatic with respect to other energy scales. This is feasible, as $\Delta E_G$ decreases exponentially with system size (Fig.~\ref{ED-fig:Fig2}(f)). 

However, with increasing system size, the extensive energy gap, $\Delta E_E$, between the polarized states and the next higher-lying states decreases linearly and closes at a critical size $N_c$ (see Fig.~\ref{ED-fig:Fig2}(e)), ultimately leading to the breakdown of quantization. This opens a window of frequencies at which there is fractional pumping: $\Omega \gg \Delta E_G$, such that a single flavor of multi-band Wannier states is followed, and $\Omega$ sufficiently small to avoid excitations. In this regime -- which we call intermediate adiabaticity --, the wavefunction follows one of the $2T$ periodic multi-band Wannier states (e.g., the orange line in Fig.~\ref{ED-fig:Fig2}), resulting in fractional pumping.

\begin{figure}[t]
    \includegraphics[width=1.0\columnwidth]{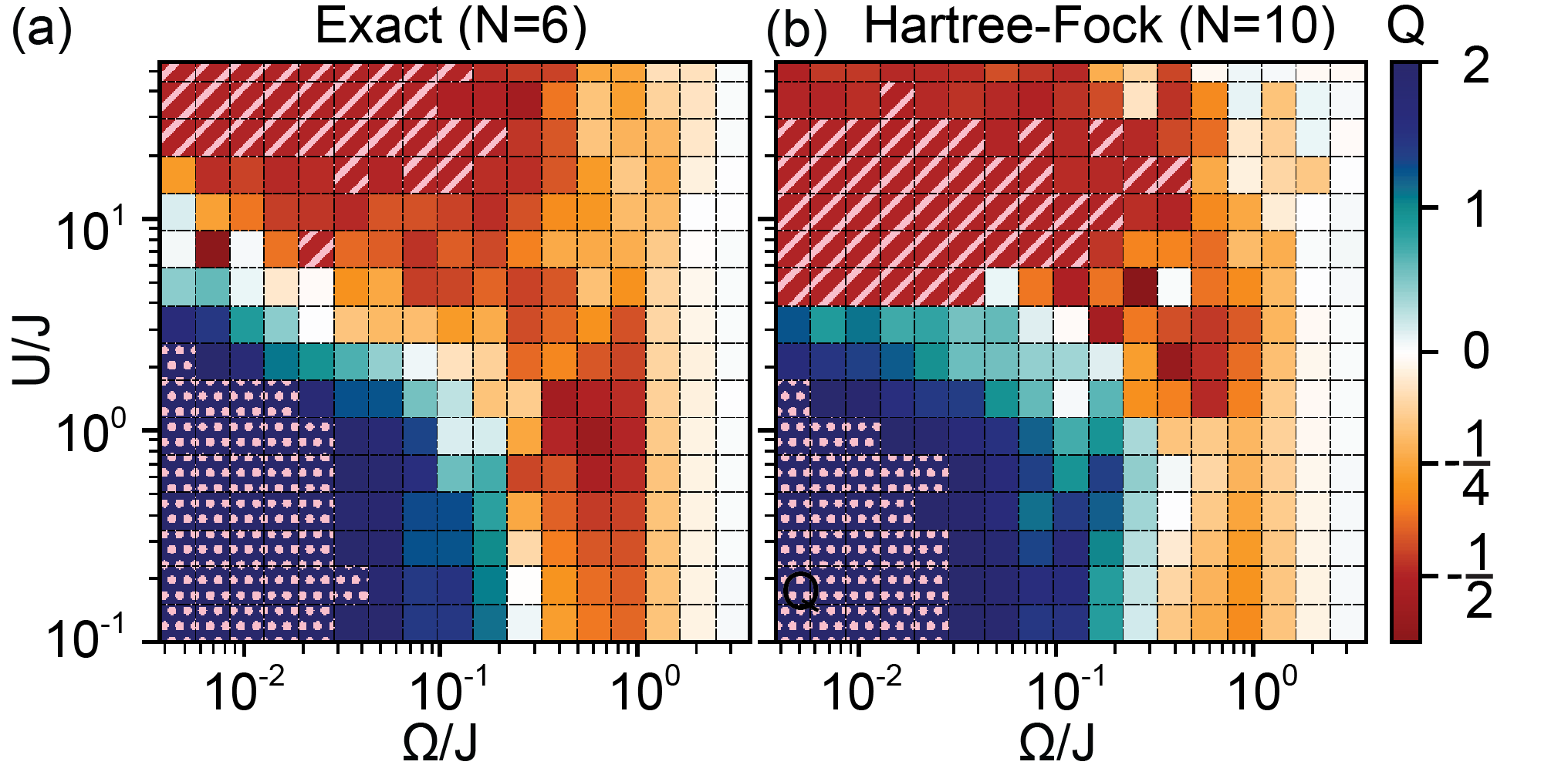}
    \caption{Phase diagrams of the interacting AAH model, obtained by exact time evolution (a) and for a self-consistent time evolution using a Hartree-Fock ansatz (b). The color scale is nonlinear and shows the pumped charge per period, $Q$, obtained as average over two periods (see SI for linear color scale). Hatched (dotted) area is within $1$\% of $Q = -0.5$ ($Q = 2.0$). Parameters: $U_r = U/r$, $1\leq r\leq 3$, and $K/J = -1.5$.}
    \label{fig:phase_diags}
\end{figure}

We confirm this by numerically solving the time-dependent Schrödinger equation $i \frac{d}{dt } \ket{\Psi(t)} = H(t)\ket{\Psi(t)}$, initializing the system in its ground state at $t=0$. We calculate the pumped charge per period over two periods via $Q = \frac{1}{2} \int_0^{2T} dt \, i\bra{\Psi(t)} J^{}_j(t) \hat{c}_{j}^{\dagger} \hat{c}^{}_{j+1} - J^{}_j(t) \hat{c}_{j+1}^{\dagger} \hat{c}^{}_{j} \ket{\Psi(t)}$.
Results using $K/J=-1.5$ are shown in Fig.~\ref{fig:phase_diags}(a), displaying the pumped charge as a function of interaction strength $U$ (with $U_r = U/r$, $1\leq r\leq 3$) and pump frequency $\Omega$. At low pumping frequencies, we observe integer pumping at weak interactions ($Q=+2$; dotted) and fractional pumping at strong interactions ($Q=-1/2$; hatched), separated by a non-quantized region. Additional data (line cuts, different system sizes and band parameters) can be found in the supplemental material \cite{SI}. These results confirm that few-particle fractional pumping indeed persists over a finite range of frequencies and interaction strengths.

\paragraph*{Toy model.}
\label{sec:multi-band_wannier_picture}
We now introduce a simplified model that captures the essential features of fractional Thouless pumping. First, we express $H_0$ in the multi-band Wannier basis (see also Fig.~\ref{ED-fig:model}(e)) keeping only the dominant terms: 
\begin{multline}
    \tilde{H}_0(t)  =  \tilde{V}(t)  \sum_{R} \bqty{ \hat{n}_{R,A}(t) - \hat{n}_{R,B}(t)  } \\ - \tilde{J}(t) \sum_{\substack{ \text{n.} \text{n.}}} \hat{w}_{R,\alpha}^{\dagger}  (t)\hat{w}^{}_{R',\overline{\alpha}}(t).
    \label{ED-Eq:WannierSpace}
\end{multline}
The second sum is restricted to nearest neighbors (n.n.), which are always of opposite flavor, but which may lie in the same unit cell $R$. $\tilde{V}$ and $\tilde{J}$ denote on-site potentials and nearest-neighbor hoppings, respectively (see \cite{SI} for definitions). Interachange of flavors after one period implies $\tilde{V}(t+T) = - \tilde{V}(t)$ \footnote{We note that $\tilde{H}_0(t) = \tilde{H}_0 (t+T)$ as the sign change of $\tilde{V}$ is compensated by the sign change of $\hat{n}_{R,A} - \hat{n}_{R,B}$.}.  
Similarly, keeping only the dominant interaction matrix elements yields: 
\begin{equation}
    \tilde{H}_{\text{int}} (t) = \tilde{U} (t) \sum_{\substack{ \text{n.} \text{n.}}}    \hat{n}_{A,R} (t) \hat{n}_{B,R'}(t),
\end{equation}
with effective interaction strength $\tilde{U}$ between nearest-neighbor Wannier states. We neglect differences between intra- and inter-unit cell hoppings and interactions.

Assuming strong interactions ($\tilde{U} \gg \tilde{V}, \tilde{J}$), we treat $\tilde{H}_0$ as a perturbation to $\tilde{H}_{\text{int}}$ (omitting time arguments for brevity). To zeroth order, the ground states are $\ket{\Psi_A}$ and $\ket{\Psi_B}$. Including the on-site potential, their energies are $E_A = N \tilde{V}$ and $E_B = -N \tilde{V}$. At the degenerate points ($\tilde{V}(t_0) = 0$) the hopping $\tilde{J}$ hybridizes $\ket{\Psi_A}$ and $\ket{\Psi_B}$. However, as $\ket{\Psi_A}$ and $\ket{\Psi_B}$ differ in every unit cell, this requires a process of order $N$, resulting in a gap that scales as $\Delta E_G = \abs{E_A(t_0) - E_B(t_0)} \sim c(N) \tilde{J}^N/\tilde{U}^{N-1} \simeq \tilde{J} e^{-\ln(\tilde{U}/4\tilde{J})N}$, confirming the numerically observed exponential scaling (with, $c(N) \sim 4^N/\sqrt{N}$ a combinatorial factor \cite{SI}). Low-energy excitations above $\ket{\Psi_A}$ and $\ket{\Psi_B}$ are domains of opposite flavor. Without loss of generality consider $\Delta > 0$. The lowest excitation, a single unit cell with flipped flavor, has energy $E_E = (-N+2) \tilde{V} + \tilde{U}$. Then, the gap to excited states is $\Delta E_E = E_E - E_B = \tilde{U} - 2 \tilde{V} (N - 1)$, decreasing linearly with $N$ as observed in the previous section. Thus, the states $\ket{\Psi_A}$ and $\ket{\Psi_B}$ are isolated from domain-like excitations only if $N < N_c \simeq \tilde{U}/(2\tilde{V}_\textrm{max})$, where $ \tilde{V}_\textrm{max} = \max_t \tilde{V}(t)$. For such systems, we expect fractionally quantized pumping up to corrections from Landau-Zener transitions between the polarized states. This correction scales as $\delta Q \sim \tilde{J}/(N \sqrt{\Omega \tilde{V}_\textrm{max}}) \exp\Bqty*{-\ln\pqty*{\tilde{U}/4\tilde{J}}N }$ \cite{SI}. The correction arises solely from the amplitdue transfer between $\ket{\Psi_A}$ and $\ket{\Psi_B}$ during the pump cycle; a superposition $a \ket{\Psi_A} + b \ket{\Psi_B}$ with $a,b$ time-independent, also yields fractional pumping. Thus for $N<N_c$ pumping remains fractionally quantized up to exponentially small corrections.

For $N > N_c$ the polarized state encounters domain excitations during the pump cycle. The smallest such excitation has length $\ell_\textrm{min} = U/(2\tilde{V}_\textrm{max}) \gg 1$ and is produced at a rate that is exponentially suppressed in $\ell_\textrm{min}$ \cite{RutkevichDecay1999,LagneseFalseVacuum2021}. We expect fractional pumping to persist until a significant number of such domains has been produced, defining a prethermal time scale. A quantitative theory of the lifetime of this prethermal fractional Thouless pump is beyond the scope of this work.

\paragraph*{Mean field theory.}

The simple picture discussed above holds only in the strong-interaction limit. For intermediate interactions, we gain insight using a mean-field approach (see \cite{SI} for details). We decouple the interaction term as
\begin{align}
    \tilde{H}_{\text{int}} (t) \to \tilde{H}_{\text{mf}}(t) =&\  \delta \tilde{V}(t)    \sum_{R} \bqty{ \hat{n}_{R,A}(t) - \hat{n}_{R,B} (t) },\\
    \delta \tilde{V} (t) =&\ 2\tilde{U} (t)\ev{\hat{n}_{R,B} (t)- \hat{n}_{R,A} (t)}.
\end{align}
The instantaneous self-consistent field $\delta \tilde{V}(t)$ effectively shifts $\tilde{V}(t) \to \tilde{V} (t)+ \delta \tilde{V}(t)$, further lowering the energy of the dominant flavor. We introduce an instructive analogy: at $\tilde{V} = 0$, the two flavors are equivalent, and $\delta \tilde{V}$ is the order parameter associated with spontaneous polarization of the flavors, akin to the magnetization of an Ising magnet. $\tilde{V}$ favors one of the polarizations, and hence acts like the magnetic field. For small $\tilde{U}$ (corresponding to the paramagnetic phase), there is only one mean-field solution $\delta\tilde{V} \propto \tilde{V}$. Thus, $\tilde{H}_{\text{mf}}(t)$ is $T$-periodic with integer pumping. For large $\tilde{U}$ (corresponding to the ferromagnetic phase), two stable solutions emerge, representing the two polarized states with $\delta\tilde{V} > 0$ and $\delta\tilde{V} < 0$. In the presence of the symmetry-breaking field $\tilde{V}$, one of these solutions becomes metastable, and vanishes if $\tilde{V}/\tilde{U}$ exceeds a threshold. However, $\tilde{V}(t)$ remains bounded throughout the pump cycle. Hence, for sufficiently strong interactions, both solutions persist throughout the cycle. Neglecting the production of domain excitations, the system adiabatically follows a single solution yielding $\tilde{H}_{\text{mf}}(t) = \tilde{H}_{\text{mf}}(t+2T)$. This period doubling, combined with $\ket{\Psi_A(t)}=\ket{\Psi_B(t+T)}$, implies fractional pumping. For intermediate interactions, the symmetry-breaking field $\tilde{V}(t)$ eliminates metastable minima during parts of the cycle, preventing adiabaitic following of the mean-field solution. This results in non-quantized pumping, and shows that the transition between integer and fractional pumping is not direct.

To confirm that the mean-field picture captures the essential features of the fractional pump we calculate the pumped charge using mean field time evolution \cite{LiuNonlinear2003,SI}. To account for essential interband processes governing the motion of Wannier states, we employ a Hartree-Fock approximation to the full model $H(t)$. As shown in Fig.~\ref{fig:phase_diags}(b), key features of the pumping phase diagram are reproduced. The main differences from the exact result are an earlier onset of the fractional pumping phase and its breakdown at large interaction strengths, which we attribute to the coupling to higher bands. This suggest that Hartree-Fock overestimates interaction effects in the present situation.

We note that the mean-field theory makes no assumption on system size. Although it neglects domain formation and other correlations, we expect it to capture short-time dynamics (few pump cycles) even for $N>N_c$.

\paragraph*{Discussion.}
We have demonstrated fractional charge pumping in interacting fermionic Thouless pumps. For suitable finite-range interactions exceeding the band gap, multi-band Wannier states are spontaneously occupied to better localize the fermions. Possible realizations include cold-atomic systems \cite{nakajimaTopologicalThoulessPumping2016,walterQuantizationItsBreakdown2023,viebahnInteractions2024} with long-range interactions, e.g. via Rydberg dressing. The fractional winding of multi-band Wannier states implies fractional pumping if the polarized state is dynamically followed. We focused on small systems where this is possible up to corrections exponentially small in system size. A quantitatively study of prethermal fractional pumping in large systems is left for future work. Notably, long-range interactions may stabilize fractional pumping even beyond prethermal times \cite{defenu2023long}.
In contrast to previous proposals \cite{grusdtRealizationFractionalChern2014,zengFractionalChargePumping2016,taddiaTopologicalFractionalPumping2017,gawatz2022prethermalization,esinUniversalTransport2024}, our mechanism does not rely on spontaneous breaking of lattice translation symmetry. Instead, fractional pumping occurs at integer band fillings, with an integer quantized phase at weak interactions. Both approaches are related to the thin-torus limit of the two-dimensional fractional quantum Hall effect \cite{taoFractionalQuantizationHall1983,rezayiLaughlinStateStretched1994,bergholtzQuantumHallSystem2008}: in our case, the role of the partially filled Landau level is played by the composite band $1+2$. Our work bridges the recently observed fractional Thouless pumping of attractive bosons (via soliton formation) and repulsive fermions, both rooted on the fractional winding of multi-band Wannier states.

\begin{acknowledgments}
We acknowledge the support of the AFOSR MURI program, under agreement number FA9550-22-1-0339 as well as the ONR MURI program, under agreement number N00014-20-1-2325.
\end{acknowledgments}

\bibliography{bibliography}

\end{document}